\documentclass[english,11pt,oneside]{article}
\pdfoutput=1
\usepackage{amsmath}
\usepackage{amsfonts}
\usepackage{amssymb}
\usepackage{graphicx, rotating}
\usepackage{epstopdf}
\usepackage{epsfig}
\usepackage{latexsym}
\usepackage{multirow}
\usepackage{color}
\usepackage{slashed}
\usepackage{pifont}
\usepackage[top=2.8 cm, bottom=2.8 cm, left=3 cm, right=3 cm]{geometry}
\usepackage[compat=1.0.0]{tikz-feynman}

\usepackage[nottoc]{tocbibind}
\usepackage{cite}

\newcommand{\ba}{\begin{eqnarray}}
\newcommand{\ea}{\end{eqnarray}}
\newcommand{\no}{\nonumber}

\usepackage{hyperref}
\hypersetup{
 colorlinks = true,
 linkcolor  = red,
 citecolor={blue}
}

\begin{document}

\title{Super-Soft CP Violation
}
\author{Alessandro Valenti$^{1\,2}$~\footnote{\href{mailto:alessandro.valenti@pd.infn.it}{\color{black}{alessandro.valenti@pd.infn.it}}}~~~~Luca Vecchi$^2$~\footnote{\href{mailto:luca.vecchi@pd.infn.it}{\color{black}{luca.vecchi@pd.infn.it}}}\\
	{$^1$\small\emph{Dipartimento di Fisica e Astronomia ``G. Galilei", Universit\'a di Padova, Italy}}\\
	{$^2$\small\emph{Istituto Nazionale di Fisica Nucleare (INFN), Sezione di Padova, Italy}}}
\date{}
\maketitle

\begin{abstract}

Solutions of the Strong CP Problem based on the spontaneous breaking of CP must feature a non-generic structure and simultaneously explain a coincidence between a priori unrelated CP-even and CP-odd mass scales. We show that these properties can emerge from gauge invariance and a CP-conserving, but otherwise generic, physics at the Planck scale. In our scenarios no fundamental scalar is introduced beyond the Standard Model Higgs doublet, and CP is broken at naturally small scales by a confining non-abelian dynamics. This approach is remarkably predictive: robustness against uncontrollable UV corrections to the QCD topological angle requires one or more families of vector-like quarks below a few $10$'s of TeV, hence potentially accessible at colliders. Because CP violation is communicated to the SM at these super-soft scales, our solution of the Strong CP Problem is not spoiled by the presence of heavy new states motivated by other puzzles in physics beyond the Standard Model. In addition, these models generically predict a dark sector that may lead to interesting cosmological signatures.

\end{abstract}

\newpage

{
	\hypersetup{linkcolor=black}
}

\section{Motivations}
\label{sec:intro}

Among the various approaches to the Strong CP Problem, scenarios that assume an {\emph{exact}} CP symmetry in the UV have a notable advantage over solutions relying on an anomalous Peccei-Quinn symmetry: this assumption can be argued to be robust against quantum gravitational effects. In particular, CP can be embedded into a gauge symmetry in extra-dimensional theories of gravity, including string theory, and remain unbroken after compactification down to 4-dimensions \cite{Choi:1992xp}\cite{Dine:1992ya}. The issue of the {\emph{quality}} of the underlying symmetry is therefore obliterated once and for all.~\footnote{The authors of \cite{Carpenter:2009zs} \cite{Dine:2015jga} argued that the possibility that CP remains unbroken down to 4-dimensions might be quite unlikely, from a multiverse perspective, at least in the context of flux compactifications. Because no general conclusion can be reached in the absence of a complete understanding of string theory compactifications, for us the important point is that the premise of this work is a perfectly defendable one: the existence of an {\emph{exact CP symmetry}} is motivated by compelling theories of gravity and {\emph{CP can generically}} remain unbroken down to 4-dimensions.}

Yet, CP must eventually be spontaneously broken in order to reproduce the Standard Model (SM) at low energies. This necessary step may introduce a sensitivity to unknown physics at high scales that can potentially jeopardize the solution of the Strong CP Problem. To avoid this the breaking must be sufficiently soft. That is, it is necessary to make sure that the CP-odd scalars $\Sigma$ responsible for the spontaneous breaking of CP have vacuum expectation values much smaller compared to the UV cutoff:
\ba\label{naturalness}
|\langle\Sigma\rangle|\ll f_{\rm UV}.
\ea 
Very explicitly, this is necessary to prevent uncontrollable higher-dimensional operators suppressed by the cutoff $f_{\rm UV}$, e.g. $c_{ab}\Sigma^\dagger_a\Sigma_b\epsilon^{\mu\nu\alpha\beta} G_{\mu\nu}G_{\alpha\beta} g_s^2/(16\pi^2f_{\rm UV}^2)$, to spoil the solution. The existence of such a hierarchy of scales \eqref{naturalness} has to be explained, otherwise no solution of the Strong CP Problem is offered. Indeed, if no justification of the hierarchy is provided the QCD theta angle would be incalculable, because susceptible of corrections from unknown corrections from the UV, and its smallness would merely be the consequence of a hidden assumption. The requirement \eqref{naturalness} may be successfully addressed by Supersymmetry~\cite{Hiller:2001qg} or a strong dynamics~\cite{Vecchi:2014hpa}. In this paper we will analyze a class of solutions of the latter type, in which CP is dynamical broken at naturally low scales.

Once \eqref{naturalness} is explained, scenarios relying on an exact CP have to address another important challenge. Spontaneous CP breaking should be properly communicated to the SM so as to guarantee that a sizable CKM phase is generated and a tiny topological angle is predicted. One way to approach this challenge was proposed in \cite{Nelson:1983zb} and \cite{Barr:1984qx}. Within ordinary 4-dimensional theories Nelson-Barr scenarios in fact represent a virtually unique option~\cite{Vecchi:2014hpa}. From an effective field theory perspective these models are characterized by a non-generic structure resulting from the selection rules associated to a global $U(1)_A$. The light degrees of freedom include the SM fields and colored fermionic messengers $\psi,\psi^c$ with the same (though vector-like) representation of the d-quarks under the SM gauge symmetry $SU(3)_c\times SU(2)_w\times U(1)_Y$. These couple to the CP-odd scalars via the operator ${\cal L}\supset y\Sigma \psi d$. Models with up-quark type mediation can also be constructed but will not be considered here because they necessitate of some additional assumption about the flavor structure of the Yukawa couplings. 

Below the scale of spontaneous breaking CP violation the scalars decouple and their effect is parametrized solely by a CP-odd mass matrix $\xi\equiv y\langle\Sigma\rangle$. The global $U(1)_A$ is realized as a spurionic symmetry under which the SM is neutral whereas the messengers, their CP-even mass $m_\psi$, and $\xi$ transform as in Table \ref{tab:EFT}.

\begin{table}[h!]
	\begin{center}
		\begin{tabular}{c||c|c|c||c}
			& $SU(3)_c$ & $SU(2)_w$ & $U(1)_Y$ & $U(1)_A$    \\
			\hline\hline
			$\psi$  & ${\bf 3}$ & ${\bf 1}$ & $-{\frac{1}{3}}$ & $z_\psi$  \\
			$\psi^c$  & $\overline{\bf 3}$ & ${\bf 1}$ & $+{\frac{1}{3}}$ & $z_{\psi^c}$  \\		
			$m_\psi$  & ${\bf 1}$ & ${\bf 1}$ & ${0}$ & $-z_{\psi}-z_{\psi^c}$  \\	
			$\xi,\Sigma$  & ${\bf 1}$ & ${\bf 1}$ & ${0}$ & $-z_{\psi}$  \\	
		\end{tabular}
		\caption{Field content beyond the SM in the minimal Nelson-Barr model. All fields are Weyl fermions. The $U(1)_A$ symmetry is global.
		}\label{tab:EFT}
	\end{center}
\end{table}

The $U(1)_A$ restricts the couplings of the messengers and the CP-odd spurion in the effective theory. In particular, it forbids CP-odd couplings between the messengers and the quark doublets which would otherwise lead to large radiative corrections to the theta angle~\cite{Vecchi:2014hpa}. The unique renormalizable couplings allowed by $U(1)_A$ and involving $\xi$ and/or the messengers, besides the standard kinetic terms, are a CP-odd mass mixing with the d-quarks and a CP-even Dirac mass: ${\cal L}\supset\xi\psi d+m_\psi\psi\psi^c+{\rm hc}$. The basic observation of \cite{Nelson:1983zb}\cite{Barr:1984qx} is that within such a framework the QCD topological angle $\bar\theta$ is radiatively generated whereas the CP-violating mass mixing generates a sizable CKM phase already at tree-level as long as $|\xi|\gtrsim m_\psi$. The radiative corrections to $\bar\theta$ are well below the current bounds if the couplings to the CP-violating sector are sufficiently small, i.e.
\ba\label{ysmall}
|y|\ll1. 
\ea
However these requirements are not enough. In fact, it turns out that in the limit ${|\xi|}\gg{m_\psi}$ the low energy theory contains a light d-quark with suppressed Yukawa couplings. The SM can thus be reproduced within a perturbative framework {\emph{only}} if~\cite{Vecchi:2014hpa}\cite{Valenti:2021rdu}
\ba\label{coincidence}
1\lesssim\frac{|\xi|}{m_\psi}\ll10^3.
\ea
This coincidence of scales cannot be explained by the effective field theory of Table \ref{tab:EFT}. It must be the consequence of some property of the UV completion. The coincidence \eqref{coincidence} is especially remarkable because $\xi$ is CP-odd whereas $m_\psi$ must be CP-even. How is it possible that quantities with an a priori qualitatively different UV origin, like the CP-even $m_\psi,y$ and the vacuum expectation value of a separate sector, turn out to be comparable to each other in size? ``Explaining" \eqref{coincidence} is tantamount to finding a class of UV completions for the effective field theory described above that naturally accommodates it. This is as key as \eqref{naturalness} if one wishes to truly solve the Strong CP Problem via spontaneous CP violation. In the case no such UV completions can be found one would have to conclude that Nelson-Barr scenarios are just a very elaborated way to trade the smallness of $\bar\theta$ with a subtle fine-tuning of the parameters.

From a model-building viewpoint, the real challenge in constructing realistic Nelson-Barr scenarios is therefore making sure that \eqref{coincidence} is realized within a framework that also explains \eqref{naturalness} and \eqref{ysmall}. Despite its fundamental importance, however, we know of no previous attempts to explain \eqref{coincidence} within familiar 4-dimensional field theories.~\footnote{This coincidence is explained by the extra-dimensional scenarios of Ref. \cite{Vecchi:2014hpa}, which would correspond to (conjectured) strongly-coupled 4-dimensional theories.} One might naively argue that \eqref{coincidence} would be easily accommodated in a theory where all couplings are of order unity and all scales beyond the SM of comparable size. A more careful look, however, reveals that this cannot be the case. In non-SUSY versions additional corrections to $\bar\theta$ always arise from loops involving excitations of the CP-violating sector and would be unacceptably large for generic couplings of order unity, see \eqref{ysmall}. But, if $y$ is taken to be small, why is the potential of the CP-violating sector minimized at a scale $\langle\Sigma\rangle\sim m_\psi/y$ that ``knows" about the couplings of the mediators to the SM? In SUSY versions of Nelson-Barr one can avoid the corrections controlled by $y$ if CP violation takes place at scales much larger than SUSY-breaking~\cite{Hiller:2001qg}. Yet, then there are necessarily several scales involved and the question remains: why would \eqref{coincidence} be satisfied? One may alternatively justify \eqref{coincidence} postulating that $m_\psi$ itself arises from the vacuum expectation value of a new scalar field, $m_\psi=y_S\langle S\rangle$. This would be an interesting approach in both non-SUSY as well as SUSY realizations. Now, granting the reasonable assumption $y\sim y_S$, the coincidence would be explained by making sure that both scalars acquire comparable vacuum expectation values, i.e. for $\langle S\rangle\sim\langle\Sigma\rangle$. The conceptual hurdle then is achieving this in such a way that $m_\psi$ be a CP-even parameter, which is mandatory if the strong CP is to be solved. Is there a way to guarantee this? Overall, we believe that the model-building implications of \eqref{coincidence} have not been fully appreciated so far. The present work will fill this gap.

In this paper we show that it is possible to find elegant UV completions of the Nelson-Barr scenarios that simultaneously address the key requirements \eqref{naturalness}, \eqref{ysmall}, and \eqref{coincidence}. In Section \ref{sec:new} we demonstrate that \eqref{naturalness}, \eqref{ysmall}, \eqref{coincidence} are naturally explained in scenarios in which CP-violation is dynamical and responsible for generating both $\xi$ and $m_\psi$, the messengers are chiral (that is they have $z_{\psi^c}\neq-z_{\psi}$ in Table \ref{tab:EFT}), and the global $U(1)_A$ is gauged. In Section \ref{sec:model} we analyze in detail an explicit realization. At low energy our UV completions reproduce the scenarios of d-mediation analyzed in \cite{Valenti:2021rdu}. This guarantees the Strong CP Problem is robustly solved. In addition, though, our constructions add new constraints and phenomenological signatures which we discuss in Section \ref{sec:pheno}. Overall, our main message is that realistic scenarios of spontaneous CP violation are very predictive and compelling solutions of the Strong CP Problem.

\section{Mediation of Super-Soft CP Breaking}
\label{sec:new}

In this section we present an elegant way to build scenarios that satisfy all the requirements reviewed in Section \ref{sec:intro}. An explicit realization will be discussed in the subsequent section. We are interested in constructing non-Supersymmetric models, continuing and completing the ideas initiated in the Appendix of \cite{Vecchi:2014hpa}.

\subsection{The Basic Setup}
\label{sec:idea}

We begin our analysis with a preliminary discussion of the key ingredients. Throughout the paper we will assume that CP is an exact symmetry in the UV. This means that there exists a field basis in which all couplings are real and the topological angles vanish. CP is then spontaneously broken in the effective field theory, as described below. We will focus on scenarios with d-mediation.

As anticipated, we tackle the hierarchy problem \eqref{naturalness} within non-Supersymmetric models. This is achieved replacing the vacuum expectation value of the fundamental scalars $\Sigma$ by the condensate of a set of SM-neutral fermion bilinears 
\ba\label{condensate}
\langle \chi_\alpha\chi_\beta^c\rangle\sim4\pi f^3\delta_{\alpha\beta}. 
\ea
Here $\chi_\alpha,\chi_\alpha^c$ are two or more families of fermions charged under, say, the fundamental and anti-fundamental of an exotic strong $SU(n)$ ($\alpha,\beta$ are the flavor indices, and to save typing we will often omit them).~\footnote{Later on we will choose $n=3$, so in this section we do not keep track of the large $n$ counting.} The latter dynamics undergoes dynamical chiral symmetry breaking at a mass scale $\sim4\pi f$, roughly the equivalent of $\Lambda_{\rm QCD}\sim1$ GeV in real-world QCD. The powers of $4\pi$ in \eqref{condensate} are borrowed from naive-dimensional analysis arguments analogous to those adopted in QCD. 

In the chiral limit the condensate does not violate CP. To see this observe that the results of~\cite{Vafa:1983tf} imply that in our scenarios the vector-like symmetries as well as parity remain unbroken, and hence $\langle \chi_\alpha\chi_\beta^c\rangle=C\delta_{\alpha\beta}$ with $C^\dagger=C$ a real number by P-invariance. Because CP acts as $C\to C^*$ it follows that the condensate is also CP invariant. The situation is completely different when the chiral symmetry is explicitly broken by tiny effects, such as higher-dimensional operators like $(\chi\chi^c)^2$, since in that case some of the approximate Nambu-Goldstone bosons emerging from chiral symmetry breaking may acquire a vacuum expectation value $\sim f$ and break CP (and/or P) spontaneously. We will later show that in our models $\langle \chi\chi^c\rangle$ generically has large CP-odd entries, with magnitude \eqref{condensate}, even if all the parameters of the Lagrangian are CP-even by hypothesis. This elegant mechanism of spontaneous CP violation of course requires at least two families of $\chi,\chi^c$, since there would be no Nambu-Goldstones otherwise. The identification $\Sigma\to \chi\chi^c$ obviously implies that the Yukawa couplings $y$ of Section \ref{sec:intro} should be replaced by non-renormalizable interactions:
\ba\label{yNonRen}
y\psi \Sigma d\to\frac{\psi d \chi\chi^c}{f_{\rm UV}^2},
\ea
for some high UV cutoff scale $f_{\rm UV}$.

The above basic setup accomplishes two goals at once. First, it ensures that the hierarchy $f\ll f_{\rm UV}$ is naturally explained via dimensional transmutation. In other words, our constraint \eqref{naturalness} is satisfied. If we are careful enough, this means we do not have to worry about possible UV effects spoiling our solution of the Strong CP Problem. Second, in a picture where \eqref{yNonRen} controls the main interaction between the SM quarks and the CP-violating sector, the non-renormalizable nature of \eqref{yNonRen} automatically guarantees that the excitations of the CP-violating sector, the hadrons of the $\chi,\chi^c$ dynamics, have very tiny couplings of order
\ba\label{yEFT}
y\sim4\pi\frac{f^2}{f_{\rm UV}^2}\ll1
\ea
with $d$ and $\psi$. Hence, potentially sizable loop corrections to the theta parameter due to the CP-violating sector are completely negligible. We have thus automatically satisfied \eqref{ysmall} as well. So far, so good. The first serious model-building challenge is, as anticipated, making sure that the coincidence \eqref{coincidence} is explained. We proceed as in the second alternative mentioned in the introduction. That is, we look for a model in which the mass of $\psi,\psi^c$ is given by a new coupling $y_S\sim y$ times the vacuum expectation value of a (composite) scalar $\langle S\rangle$, of the same order as $\langle \Sigma\rangle$. 

Since in the present framework $y$ arises from a non-renormalizable operator, also the mass of $\psi,\psi^c$ has to arise from a dimension-6 operator similar to \eqref{yNonRen}. We therefore introduce another composite scalar made up of new SM-neutral fermions with $SU(n)$ charges, i.e. $S\to\lambda\lambda$, and look for models in which the explicit mass term of Table \ref{tab:EFT} is UV-completed by
\ba\label{mNonRen}
m_\psi\psi\psi^c\to\frac{\psi\psi^c\lambda\lambda}{f_{\rm UV}^2}.
\ea
As long as the UV dynamics is sufficiently generic (but of course CP-invariant), the coefficients in front of \eqref{yNonRen} and \eqref{mNonRen} should be comparable in size, implying $y\sim y_S$ as needed. In addition, $|\langle\lambda\lambda\rangle|\sim|\langle\chi\chi^c\rangle|$ would be generically satisfied because the $\lambda$'s are charged under the very same confining $SU(n)$ carried by $\psi,\psi^c$. With these assumptions we get
\ba\label{mNonRen1}
m_\psi\sim\frac{\langle\lambda\lambda\rangle}{f_{\rm UV}^2}\sim\frac{|\langle\chi\chi^c\rangle|}{f_{\rm UV}^2}\sim |y\langle\Sigma\rangle|.
\ea
This is the desired result \eqref{coincidence}. We are making progress, but the main hurdle comes next: why does ${\rm Arg}[m_\psi]$ vanish, that is why is $\langle\lambda\lambda\rangle$ real while $\langle\chi\chi^c\rangle$ is complex?

Since $\psi,\psi^c$ are colored, an hypothetical phase in their mass would immediately translate into a correction to $\bar\theta$. If no further assumption is made, $\langle\lambda\lambda\rangle$ is expected to carry its own broken chiral symmetries and be complex, as we argued for $\langle\chi\chi^c\rangle$. Fortunately this can be avoided. We identified four sufficient conditions that guarantee that the complex phases in $\langle\lambda\lambda\rangle$ are sufficiently small to be compatible with $|\bar\theta|\lesssim10^{-10}$. These are:
\begin{itemize}
\item[(a)] $\lambda$ must appear in a single family. 
\item[(b)] There must be a gauge $U(1)$ under which the $SU(n)$ sector is chiral. 
\item[(c)] The gauge $U(1)$ must commute with the non-abelian flavor symmetry of the $\chi,\chi^c$'s. 
\item[(d)] The scale of spontaneous CP violation has to satisfy
\ba\label{epsilonBound}
\frac{f}{f_{\rm UV}}\lesssim10^{-5}.
\ea
\end{itemize}
We will see in the next subsection how our conditions can be implemented in concrete models. Here we explain why $\langle\lambda\lambda\rangle$ is approximately CP-even when they are satisfied. 

To start, (a) implies $\lambda$ does not carry non-abelian global symmetries which would otherwise generically imply large complex phases arise from the vacuum expectation value of the associated Nambu-Goldstone fields. The unique spontaneously broken global symmetry $\lambda$ is allowed to carry is an {\emph{axial abelian}} one, if present. In fact, in Section~\ref{sec:intro} we argued that such a symmetry must be there in order to reproduce the desired structure. It should not be a surprise to find then that the global charge carried by $\lambda$ must be precisely the $U(1)_A$ of Table \ref{tab:EFT}. To see this assume that such a $U(1)_A$ exists. It follows that the new strong sector has to be charged under it if we want to allow \eqref{yNonRen}. In particular, $\chi\chi^c$ has to be chiral under the $U(1)_A$. From this observation one might directly conclude that $\lambda$ must also be charged. Indeed an accurate $U(1)_A$ can only survive in the IR if such a symmetry has no $U(1)_A\times SU(n)\times SU(n)$ anomaly. This of course requires the fermion $\lambda$ to be charged. There is an alternative argument that forces $\lambda$ to be chiral under the very same $U(1)_A$. Because $m_\psi$ is to be given by \eqref{mNonRen} we better make sure that $\psi,\psi^c$ are themselves chiral under $U(1)_A$, otherwise a Dirac mass term for $\psi,\psi^c$ would be allowed and we would not be able to convincingly explain \eqref{coincidence}. The bilinear $\lambda\lambda$ has thus to be charged as well. However we want to put it, the necessary low-energy structure of these models combined with \eqref{yNonRen} and \eqref{mNonRen} lead us to conclude that the $SU(n)$ sector must be chiral under the global $U(1)_A$.

Under the hypothesis (a) we see that the only Nambu-Goldstone mode that can potentially induce a sizable complex phase in $\langle\lambda\lambda\rangle$ is the one associated to the $U(1)_A$. However, condition (b) renders the latter unphysical: the phase in $\langle\lambda\lambda\rangle$ due to the vacuum expectation value of the (would-be) $U(1)_A$ Nambu-Goldstone boson is eaten by the abelian gauge field. Importantly, for this to fully hold (c) must be satisfied. According to (c), the gauge charges of $\chi_\alpha,\chi^c_\alpha$ must be the same for each flavor $\alpha$. This ensures that the {\emph{gauge}} $U(1)$ acts on the $SU(n)$ sector exactly as the {\emph{global}} $U(1)_A$. As a consequence, the longitudinal component of the gauge boson exactly coincides with the $U(1)_A$ Nambu-Goldstone. In the unitary gauge this is removed and cannot show up in $\langle\lambda\lambda\rangle$.

Whenever (a), (b), (c) are satisfied the only Nambu-Goldstone bosons that can contribute to the phase of $\langle\lambda\lambda\rangle$ are those of the non-abelian flavor symmetry acting on $\chi_\alpha,\chi_\alpha^c$. But because $\lambda$ is neutral under such a symmetry, their effect is proportional to the small chiral symmetry-breaking couplings that are responsible for triggering spontaneous CP violation, see below \eqref{condensate}. These come from operators of at least dimension-6 in our scenarios (see next section) and therefore lead to 
\ba\label{argm}
{\rm Arg}(m_\psi)\sim\frac{f^2}{f_{\rm UV}^2}.
\ea
The experimental constraint $|\bar\theta|\lesssim10^{-10}$ becomes an interesting upper bound \eqref{epsilonBound} on the scale of CP-breaking. The vacuum expectation values of the Nambu-Goldstone modes have a completely negligible impact on $m_\psi$ if (d) is assumed.

Interestingly, \eqref{epsilonBound} also guarantees that the contamination of other CP-odd phases does not spoil our solution of the Strong CP Problem. In particular, one may fear uncontrollable complex contributions to $\langle\lambda\lambda\rangle$ (and, less relevantly, to $\langle \chi\chi^c\rangle$) arising from the vacuum expectation value of any of the CP-odd {\emph{massive}} hadrons $\eta$ of the exotic dynamics. In general the potential of the composite scalars is the sum of a zeroth order term from the renormalizable part of the $\lambda,\chi,\chi^c$ interactions, plus a small perturbation: $V=V_0+V_1$. In our models all perturbations are due to higher-dimensional operators of at least dimension six because of the chiral nature of the $U(1)$ (see next section for details), and therefore $V_1/V_0\sim f^2/f_{\rm UV}^2$. We have seen above that an hypothetical complex phase in $\langle\lambda\lambda\rangle$ must come at next to leading order, and therefore be controlled entirely by the small perturbation $V_1$. This leads us again to \eqref{argm}. The condition (d) prevents these effects from appreciably affecting the $\bar\theta$ parameter. We stress that the results of~\cite{Vafa:1983tf} are central to our conclusions. In a theory with fundamental scalars $\Sigma, S$ we would not have at our disposal such powerful theorems and it would be difficult to find general conditions guaranteeing $m_\psi= y_S\langle S\rangle$ be CP-even.

We thus have found a picture in which the key requirements \eqref{naturalness}, \eqref{ysmall}, and \eqref{coincidence} discussed in Section \ref{sec:intro} are structurally realized and the Strong CP Problem can be robustly addressed. Remarkably, this basic set up has very important phenomenological consequences. Together with \eqref{mNonRen1} and \eqref{condensate}, eq. \eqref{epsilonBound}  implies an upper bound on the messengers' mass:
\ba\label{mBound}
m_\psi\sim4\pi f\frac{f^2}{f_{\rm UV}^2}\lesssim10^{-14}f_{\rm UV}.
\ea
The scale at which CP violation is communicated to the SM, which is set by the messengers' mass, satisfies $m_\psi\ll 4\pi f\ll 4\pi f_{\rm UV}$ and is therefore {\emph{super-soft}}. We will emphasize in Section \ref{sec:discussion} that this feature has important theoretical implications. There are also interesting phenomenological consequences, however.  Since our effective field theory arguments are at most reliable up to the Planck scale we expect $f_{\rm UV}\lesssim 2.4\times10^{18}$ GeV. It follows that
\ba\label{scaleIR}
m_\psi\lesssim{\rm few~10's~TeV}.
\ea
This constraint must be interpreted at an order of magnitude level, given it depends on an unknown UV scale and the coefficients of dimension-6 interactions, as well as the incalculable value of the condensates $\langle \chi_\alpha\chi_\beta^c\rangle$, $\langle \lambda\lambda\rangle$. Yet, the implication is clear: our solutions predict new colored fermions not far from the TeV. This important constraint makes these scenarios testable and very predictive.~\footnote{An analogous connection with the TeV was made in the context of mirror-world models in \cite{Hook:2014cda}.} We may also reverse the argument and observe that, since the messengers are colored fermions, the lack of experimental evidence of such particles says that $m_\psi\gtrsim10^3$ GeV (see Section \ref{sec:collider}). Via \eqref{mBound} this implies $f_{\rm UV}\gtrsim10^{17}$ GeV. Assuming order one coefficients in \eqref{mNonRen1}, we see the UV cutoff must lie close to the Planck scale. 

The main ingredients of the model have now been identified. In the next section we will construct a concrete realization and discuss some of the main phenomenological implications.

\section{An Explicit Model}
\label{sec:model}

We now build an UV completion for the model of Table \ref{tab:EFT} following the ideas introduced in Section \ref{sec:idea}.

\subsection{The Model}

An anomaly-free realization of the ideas in Section \ref{sec:idea} requires more fields than the minimal ones necessary to address the Strong CP Problem, namely more than just $\psi,\psi^c,\chi_\alpha,\chi^c_\alpha,\lambda$. Some of the extra states may lead to interesting phenomenological signatures, which will be analyzed later on.

The particle content beyond the SM involves only fermionic (Weyl) fields and is summarized in Table \ref{table}. The non-abelian gauge groups are all asymptotically free and the Landau poles of the abelian sector are many orders of magnitude above the Planck scale. The embedding of the fields charged under the SM in complete grand-unified $SU(5)\supset SU(3)_c\times SU(2)_w\times U(1)_Y$ multiplets is straightforward (see the caption of Table \ref{table}). Let us discuss the role of the various fields in turn.

\subsubsection{Field Content}

\begin{table}[!t]
	\begin{center}
		\begin{tabular}{c||ccccc}
			& $SU(3)_c$ & $SU(2)_w$ & $U(1)_Y$ & $SU(3)$ & $U(1)$    \\
			\hline\hline
			$\psi_1$ & ${\bf 3}$ & ${\bf 1}$ & $-\frac{1}{3}$ & ${\bf 1}$ & $+1$  \\
			$\psi_2$ & ${\bf 3}$ & ${\bf 1}$ & $-\frac{1}{3}$ & ${\bf 1}$ & $-1$  \\
			$\psi_1^c$ & $\overline{\bf 3}$ & ${\bf 1}$ & $+\frac{1}{3}$ & ${\bf 1}$ & $-\frac{1}{3}$  \\
			$\psi_2^c$ & $\overline{\bf 3}$ & ${\bf 1}$ & $+\frac{1}{3}$ & ${\bf 1}$ & $+\frac{1}{3}$  \\
			\hline
			$\psi'_1$ & ${\bf 1}$ & ${\bf 2}$ & $+\frac{1}{2}$ & ${\bf 1}$ & $+1$ \\
			$\psi'_2$ & ${\bf 1}$ & ${\bf 2}$ & $+\frac{1}{2}$ & ${\bf 1}$ & $-1$ \\
			$\psi'^{c}_1 $ & ${\bf 1}$ & ${\bf 2}$ & $-\frac{1}{2}$ & ${\bf 1}$ & $-\frac{1}{3}$ \\
			$\psi'^c_2 $ & ${\bf 1}$ & ${\bf 2}$ & $-\frac{1}{2}$ & ${\bf 1}$ & $+\frac{1}{3}$ \\
			\hline
			$\chi_{\alpha=1,2}$ & ${\bf 1}$ & ${\bf 1}$ & $0$ & ${\bf 3}$ & $+\frac{1}{2}$  \\
			$\chi^c_{\alpha=1,2}$ & ${\bf 1}$ & ${\bf 1}$ & $0$ & $\overline{\bf 3}$ & $+\frac{1}{2}$  \\
			$\lambda$ & ${\bf 1}$ & ${\bf 1}$ & $0$ & ${\bf 8}$ & $-\frac{1}{3}$  \\
			\hline
			$N_{I=1,2,3,4}$ & ${\bf 1}$ & ${\bf 1}$ & $0$ & ${\bf 1}$ & $-\frac{2}{3}$  \\
			$N'_{I=1,2,3,4}$ & ${\bf 1}$ & ${\bf 1}$ & $0$ & ${\bf 1}$ & $-\frac{1}{6}$ 
		\end{tabular}
		\caption{\label{tab:s1BSMcontent}Field content beyond the SM. All fields are Weyl fermions. Note that the messengers $(\psi_a,\psi'_a)\oplus(\psi^c_a,\psi'^c_a)$ form complete ${\bf5}_a\oplus\overline{\bf5}_a$ multiplets of a grand-unified $SU(5)$ SM gauge group, but are chiral under $U(1)$.
		}\label{table}
	\end{center}
\end{table}

\paragraph{The CP-violating Sector} 

The minimal CP-violating sector realizing the program spelled out in Section \ref{sec:idea} is composed of two families of $\chi,\chi^c$ in the fundamental representation of a new confining $SU(3)$ gauge group and a single Weyl fermion $\lambda$ in the adjoint representation. These fermions are all charged under the axial gauge $U(1)$, with charges chosen such that the anomaly $SU(3)\times SU(3)\times U(1)$ is absent. 

This theory, when supplemented with small (CP-conserving by hypothesis) perturbations of the type $(\chi\chi^c)(\chi\chi^c)^\dagger/f_{\rm UV}^2$, breaks CP spontaneously. To assess the qualitative behavior of the non-perturbative dynamics let us first neglect all couplings except for the $SU(3)$ gauge interaction. Then $\chi,\chi^c,\lambda$ enjoy a global anomaly-free symmetry $SU(2)_\chi\times SU(2)_{\chi^c}\times U(1)_V\times U(1)_A$, where $U(1)_V$ is just the $\chi,\chi^c$ baryon number, whereas the global $U(1)_A$ acts on $\chi,\chi^c,\lambda$ precisely as the $U(1)$ of Table \ref{tab:s1BSMcontent}. While there is no definite proof, there are good reasons to expect that in the IR this theory develops two condensates $\langle \chi\chi^c\rangle\sim\langle\lambda\lambda\rangle$. Indeed, according to the arguments of \cite{Vafa:1983tf} the vectorial subgroup $SU(2)_{\chi+\chi^c}\times U(1)_V$ must remain intact. Hence the only allowed condensates are $\langle\lambda\lambda\rangle, \langle \chi\chi^c\rangle$. An heuristic argument based on the most attractive channel \cite{Raby:1979my} suggests that $\langle\lambda\lambda\rangle$ is likely to form ``first", immediately followed by $\langle \chi\chi^c\rangle$. In the following we will therefore assume that both condensates form, and that they have comparable sizes.

After chiral symmetry breaking the $U(1)_A$ Goldstone mode is eaten by the $U(1)$ gauge via the Higgs mechanism. The remaining 3 Nambu-Goldstone bosons $\pi^q$ are described by the $SU(2)$ matrix $U=e^{i\pi^q\sigma^q/f}$, with $\sigma^q$ the Pauli matrices. Below the chiral symmetry breaking scale, the condensates may be parametrized as 
\ba\label{condEFT}
\langle \chi_\alpha\chi_\beta^c\rangle&=&c_\chi4\pi f^3U_{\alpha\beta}\\\no
\langle\lambda\lambda\rangle&=&c_\lambda 4\pi f^3
\ea
for some unknown parameters $c_\chi, c_\lambda$ expected to be of order unity. The latter are guaranteed to be CP-even, as demonstrated in Section \ref{sec:idea}. The renormalizable theory we just described conserves CP and has a degenerate vacuum parametrized by any value of the $\pi^q$'s. However, additional small interactions can break explicitly the accidental $SU(2)_\chi\times SU(2)_{\chi^c}$ symmetry and generate a potential for $U$. For example, a set of unavoidable dimension-6 interactions of the type 
\ba\label{chichichichi}
\frac{c_{\alpha\beta;\gamma\delta}}{f^2_{\rm UV}}(\chi_\alpha \chi_\beta^c)(\chi_\gamma \chi^c_\delta)^\dagger
\ea
breaks $SU(2)_\chi\times SU(2)_{\chi^c}$ completely if the (real) coefficients $c_{\alpha\beta;\gamma\delta}$ are generic. This operator in fact describes the dominant source of explicit chiral symmetry breaking in the model of Table \ref{tab:s1BSMcontent}. Once \eqref{chichichichi} is included in the picture, the Nambu-Goldstone bosons acquire a small potential (see \eqref{condEFT}) $V_{\rm NGB}(\pi)\sim16\pi^2f^4(f/f_{\rm UV})^2\,c_{\alpha\beta;\gamma\delta}U_{\alpha\beta}U^*_{\gamma\delta}$, and the vacuum degeneracy is lifted. The actual vacuum configuration of the $U$ field depends on the unknown CP-even parameters $c_{\alpha\beta;\gamma\delta}$. What matters to us here is not the exact expression of the vacuum state, however, but simply that CP is generically broken by complex entries of order \eqref{condensate}. To prove this one has to observe that CP is spontaneously broken if $\langle U\rangle^*\neq \langle U\rangle$, i.e. if $\langle\pi_1\rangle\neq0$ or $\langle\pi_3\rangle\neq0$. It is then easy to see that, for generic CP-even coefficients $c_{\alpha\beta;\gamma\delta}$, $V_{\rm NGB}$ is indeed minimized at~\footnote{As usual, all these expressions are to be interpreted as holding in the field basis in which all couplings are CP-even, according to the hypothesis of exact CP invariance of the UV (see Section \ref{sec:intro}).}
\ba\label{NGBvev}
{\rm Im}(\langle U\rangle)\sim1. 
\ea
A numerical scan of the coefficients $c_{\alpha\beta;\gamma\delta}$ in the range $[-10,10]$ confirms that this holds in more than $50\%$ of the parameters space: spontaneous CP-violation is generic in our model.

\paragraph{The Mediator Sector}

The messengers $\psi,\psi^c$ of Table \ref{tab:s1BSMcontent} are vector-like under the SM subgroup $SU(3)_c\times U(1)_Y$, but chiral under the new gauge $U(1)$. They appear in four different representations in order to avoid an anomaly in $SU(3)_c\times SU(3)_c\times U(1)$ and $U(1)_Y\times U(1)_Y\times U(1)$. This implies the existence of a non-minimal number of mediators' families. Of course other constructions are possible, but we will stick to this one in the following.

The particles $\psi',\psi'^c$ are introduced in Table \ref{tab:s1BSMcontent} with the sole scope of removing the $U(1)_Y\times U(1)\times U(1)$ anomaly left by the $\psi,\psi^c$ system. The $\psi',\psi'^c$ mix with the SM lepton doublets similarly to how $\psi,\psi^c$ mix with quarks. In this sense they can be seen as mediators for CP-violation in the leptonic sector. Their presence does not spoil the solution of the Strong CP Problem, as will be argued below. If preferred, it is possible to find alternative scenarios where the leptonic messengers are replaced by other fields with SM charges. In order for such alternatives to be phenomenological viable, though, the new states must have sufficiently large decay rates into SM particles. This is certainly the case with the field content of Table \ref{tab:s1BSMcontent}.

\paragraph{Extra States}

The last degrees of freedom that need to be discussed are the $N,N'$ particles. This SM neutral sector removes the remaining $U(1)\times U(1)\times U(1)$ and $U(1)\times{\rm grav}\times{\rm grav}$ anomalies. One may find many different ways to cancel these anomalies, and the field content of Table \ref{tab:s1BSMcontent} is just an arbitrary choice. A different choice will be discussed in Section \ref{sec:baryogenesis}. It is worth to make a few comments, though. First, one cannot replace the spectator sector by increasing the number of $\chi,\chi^c,\lambda$ families. If this avenue is pursued, the $SU(3)$ sector would contain flavors with different $U(1)$ charges; as a consequence the spontaneously broken abelian global symmetry and the gauged $U(1)$ would no more coincide. This would be a disaster because it would imply that the phase from $\lambda\lambda$ could not be removed by the Higgs mechanism and the Strong CP Problem would not be solved. Second, we were unable to replace the SM-neutral $N,N'$ with a slightly more involved (but still unstable) messenger sector of SM-charged states. This does not mean it is not possible, of course. Overall, our opinion is that a separate SM- and $SU(3)$-neutral spectator sector is a rather generic feature of our models.

\subsubsection{Interactions}

There is a very limited set of new couplings at the renormalizable level: the obvious kinetic terms (including those of the exotic gauge fields) and, since our model contains an abelian group, a kinetic mixing between hypercharge and the $U(1)$ gauge. The latter is however not relevant phenomenologically because the exotic vector acquires a large mass, as we will see below. On the other hand, there are no renormalizable couplings among the exotic fermions of Table \ref{tab:s1BSMcontent} and the SM fermions, and the former are all chiral. Non-gauge interactions involving the exotic fermions all arise from operators of dimension six or higher suppressed by powers of the UV cutoff $f_{\rm UV}\sim10^{17-18}$ GeV, see the discussion around \eqref{epsilonBound}. Operators of dimension larger than $6$ lead to effects suppressed by at least $v/f_{\rm UV}\lesssim10^{-15}$ or $(f/f_{\rm UV})^3\lesssim10^{-15}$ and have no phenomenological impact. (There is a single exception to this conclusion, i.e. the mass of $N'$, which we discuss around \eqref{N'mass}.) In this subsection we therefore limit ourselves to a discussion of the leading dimension-6 interactions.

We are interested in 4-fermion operators $O$ involving the fermions of Table \ref{table}. These appear in the effective Lagrangian as $cO/f_{\rm UV}^2$. Our working hypothesis of exact CP invariance at the Planck scale implies that all Wilson coefficients $c$ are real in a suitable field basis, which we assume throughout the paper. For definiteness we will take them to be of order unity. No particular flavor structure is favored by our models.

Operators involving only the fermions $\psi, \psi^c, \psi', \psi'^c, N, N'$ and SM fields are strongly irrelevant. In particular they do not bring CP-violating phases and do not contribute appreciably to $\bar\theta$. On the other hand, operators containing $\chi,\chi^c,\lambda$ can become important at low energies because after symmetry breaking they can be turned into {\emph{relevant}} interactions. Let us thus focus on interactions with $\chi,\chi^c,\lambda$. The charge assignments of Table \ref{tab:s1BSMcontent} imply these belong to five distinct classes. First we find the operators that, after symmetry breaking, generate CP-violating mass terms as in \eqref{yNonRen}:
\ba\label{dim6CP}
(\chi_\alpha\chi^c_\beta)^\dagger \psi_1 d,~~~~
\chi_\alpha\chi^c_\beta \psi_2 d,~~~~
(\chi_\alpha\chi^c_\beta)^\dagger \psi'_1 \ell,~~~~
\chi_\alpha\chi^c_\beta \psi'_2 \ell.
\ea
Note that the charge assignments do not allow analogous (and dangerous) dim-6 interactions among $\chi\chi^c$, $\psi,\psi^c$ and the quark doublet $q$. In particular, the $U(1)$ charges are carefully chosen to forbid $\psi\psi^c\chi\chi^c$, which would generate a complex mass for the quark mediators and re-introduce a Strong CP problem. In the second class we have operators inducing CP-even masses of the type \eqref{mNonRen1} after symmetry breaking. These are
\ba\label{dim6m}
\psi_1\psi^c_1 \lambda\lambda,~~~~
\psi_2\psi^c_2 (\lambda\lambda)^\dagger,~~~~
\psi'_1\psi'^c_1 \lambda\lambda,~~~~
\psi'_2\psi'^c_2 (\lambda\lambda)^\dagger.
\ea
Third, we have operators involving only $\chi,\chi^c,\lambda$, like the ones in eq. \eqref{chichichichi}. These are especially important because they represent the dominant source of explicit breaking of the chiral $SU(2)_\chi\times SU(2)_{\chi^c}$ symmetry group of the strong $SU(3)$ sector. Since we have already argued that CP gets spontaneously broken we do not need to analyze them in detail here. The fourth class of operators containing $\chi,\chi^c,\lambda$ consists of  
\ba\label{decayHad}
\chi^\dagger_\alpha\bar\sigma^\mu\chi_\beta~J_\mu,~~~~\lambda^\dagger\bar\sigma^\mu\lambda~J_\mu, 
\ea
with $J_\mu$ indicating any gauge singlet vector current (of dimension 3) constructed out of the other fields, SM fields included. After chiral symmetry breaking this class of operators generate tiny (in general CP-violating) interactions between the $SU(3)$ hadrons and the rest of the world. Fortunately, they also do not affect $\bar\theta$ appreciably because they induce corrections suppressed by powers of $(f/f_{\rm UV})^2\lesssim10^{-10}$. Finally, the last class of operators involving the CP-violating sector parametrize flavor-violating interactions with the spectator sector:\ba\label{NmassMix}
\chi_\alpha\chi_\beta^c\lambda N_I.
\ea
This generates, after $SU(3)$ confinement, a mass for $N_I$, as we will discuss below. On the other hand, $N'_I$ remains massless at dimension-6. We will estimate its mass in the next subsection.

\subsection{Phenomenology}
\label{sec:pheno}

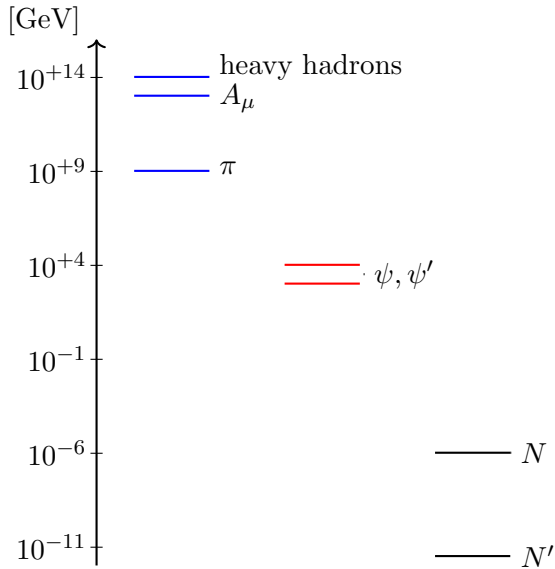
\begin{figure}[t]
	\begin{center}
		\begin{tikzpicture}[scale = 0.25, level/.style={thick}, font=\normalsize ]
		
		\draw[->, thick] (0cm, -8cm) -- (0cm, 20cm)node at (0, 21cm)[xshift =-2pt, left]{[GeV]};
		\draw[-|](0,18cm) node [left]{$10^{+14}$ };
		\draw[-|](0,13cm) node [left]{$10^{+9}$ };
		\draw[-|](0,8cm) node [left]{$10^{+4}$ };
		\draw[-|](0,3cm) node [left]{$10^{-1}$};
		\draw[-|](0,-2cm) node [left]{$10^{-6}$};
		\draw[-|](0,-7cm) node [left]{$10^{-11}$ };
		\draw[level, blue] ( 2 cm,18cm) -- ( 6cm, 18cm); \draw[] ( 6cm, 18cm) node[yshift = +4pt,right]{heavy hadrons};
		\draw[level, blue] ( 2cm, 17cm) -- ( 6cm, 17cm); \draw[] ( 6cm, 17cm) node[yshift = -1pt, right]{$A_{\mu}$};
		\draw[level, blue] ( 2cm, 13cm) -- ( 6cm, 13cm); \draw[] ( 6cm, 13cm) node[yshift = +1pt, right]{$\pi$};

		\draw[level, red] ( 10cm, 8cm) -- ( 14cm, 8cm);\draw[] ( 14cm, 8cm) node[yshift = +4pt, right]{$$};
		\draw[level, red] ( 10cm, 7cm) -- ( 14cm, 7cm);\draw[] ( 14cm, 7cm) node[yshift = +1pt, right]{$$};

		\draw[level] ( 14.2cm, 7.5cm) -- ( 14.2cm, 7.5cm) node[right]{$\psi, \psi'$};

		\draw[level] ( 18cm, -2cm) -- ( 22cm, -2cm);
		
		\draw [level] (22cm,-2cm) -- (22cm,-2cm) node [right]{$N$};
		
				
		\draw[level] ( 18cm, -7.5cm) -- ( 22cm, -7.5cm) node[right]{$N'$};
		
		\end{tikzpicture}
	\end{center}
	\caption{\label{fig:s2scales}
	Schematic representation of the mass scales involved. See the text for details.}
\end{figure}

In this subsection we analyze in some detail the phenomenology of our model. Yet, before embarking in this journey, it proves useful to summarize the different mass scales in the theory. These are pictorially shown in Fig \ref{fig:s2scales}.


The highest scale is the UV cutoff, parameterized by $f_{\rm UV}\sim10^{17-18}$ GeV. One may identify it with the Planck scale, but we decided to keep our discussion more general. All masses of the particles beyond the SM masses arise from the dynamically generated scale 
\ba
m_{\rm CP}\sim4\pi f\sim10^{13-14}~{\rm GeV}
\ea
of the confining $SU(3)$ sector. The natural hierarchy ${f}/{f_{\rm UV}}$ appears in many of the following expressions, so it is convenient to introduce the more compact notation
\ba
\epsilon\equiv\frac{f}{f_{\rm UV}}.
\ea
As we argued around eq. \eqref{epsilonBound}, $\epsilon\lesssim10^{-5}$ ensures there are no large contributions to the QCD $\bar\theta$ parameter due to the vacuum expectation value of CP-odd resonances.

Let us now see what masses arise from chiral symmetry breaking. First, heavy $SU(3)$ hadrons all have masses of order $m_{\rm CP}$. The $SU(3)$ dynamics also generates pseudo Nambu-Goldstone bosons, the key players in the spontaneous breaking of CP (see \eqref{NGBvev}). Their masses are induced dominantly by the interactions in \eqref{chichichichi} and are expected to be of order (see the corresponding potential $V_{\rm NGB}$)
\ba
m_\pi\sim m_{\rm CP}\epsilon\lesssim10^{8-9}~{\rm GeV}.
\ea
Furthermore, chiral symmetry breaking generates a mass for the $U(1)$ vector, $m_A\sim g_Af\sim(g_A/4\pi)m_{\rm CP}$, with $g_A$ the $U(1)$ gauge coupling. For definiteness we will assume that $g_A$ is not far from order unity, so that $m_A\gg m_\pi$. This assumption does not have any significant impact on our analysis, though.

Similarly, the fermions $\psi,\psi^c$ and separately $\psi',\psi'^c$ form, after chiral symmetry breaking, two families of Dirac pairs with CP-even masses generated by the interactions \eqref{dim6m} and CP-odd mixings \eqref{dim6CP} with the SM $d,\ell$ representations. Overall, these are of order $m_\psi,m_{\psi'}\sim m_{\rm CP}\epsilon^2\lesssim1-10~{\rm TeV}$ (see also \eqref{mNonRen1}). 

The spectator sector of Table \ref{table} lives at scales parametrically smaller than the TeV. $N$ gets a mass after a seesaw-like mixing with heavy fermionic hadrons $\sim\chi\chi^c\lambda$ via \eqref{NmassMix}. A rough estimate says that
\ba\label{Nmass}
m_N\sim\epsilon^4m_{\rm CP}\lesssim10^{2}-10^{3}~{\rm eV}.
\ea
Finally, the dominant contribution to the mass of $N'$ arises from dimension-9 interactions like $N'N'\chi\chi^c\lambda\lambda$. After chiral symmetry breaking we get
\ba\label{N'mass}
m_{N'}\sim m_{\rm CP}\epsilon^5\lesssim10^{-3}-10^{-2}~{\rm eV}.
\ea
Note that the larger powers of $\epsilon$ in \eqref{Nmass}, \eqref{N'mass} compared to the other beyond the SM particles make $m_{N,N'}$ more sensitive to the actual ${\cal O}(1)$ couplings involved in our estimates. The cosmological signatures of this sector will be discussed in Section \ref{sec:cosmo}. 

We next turn to a study of how the Strong CP is solved, and then cosmological and collider signatures of our scenarios.

\subsubsection{The CKM Phase and the Strong CP Problem}

It is easy to see why the model introduced in Section \ref{sec:model} solves the Strong CP Problem. CP is spontaneously violated by the vacuum expectation value $\langle \chi\chi^c\rangle$. The excitations of the CP-violating sector (heavy hadrons and Nambu-Goldstone bosons) couple to the SM via small Planck-suppressed couplings \eqref{yEFT} and can be ignored, as anticipated in \eqref{ysmall}. For all practical purposes the CP-violating sector is frozen and parametrized by the two condensates $\langle \chi\chi^c\rangle$ and $\langle \lambda\lambda\rangle$. Within the effective field theory at scales $\ll m_\pi\ll4\pi f$ the relevant degrees of freedom are the SM particles and $\psi,\psi^c,{\psi'},{\psi'}^c,N,N'$. There is a unique CP-odd spurion, namely $\langle \chi\chi^c\rangle$, which couples to the colored sector solely via \eqref{dim6CP}. The CP-violating theory we are describing is essentially that of Table \ref{tab:EFT} with $\xi$ replaced by ${\langle \chi\chi^c\rangle}/{f_{\rm UV}^2}$ and mediators' masses satisfying $|m_\psi|\sim|\xi|$ (see \eqref{mNonRen1}). This theory reproduce the SM at scales $\ll m_\psi$, including the CKM phase, and a $\bar\theta$ very comfortably below the current bounds \cite{Valenti:2021rdu}.

Yet, our UV completion adds new ingredients to the effective field theory of Table \ref{tab:EFT}. It predicts the constraint \eqref{scaleIR} and additional (SM-charged) unstable particles ${\psi'},{\psi'}^c$, neutral states $N,N'$, and Planck-suppressed CP-conserving interactions. The additional states cannot play any role in transferring CP-violation to QCD. In particular, the leptonic CP-violating coupling in \eqref{dim6CP} is completely irrelevant for the Strong CP Problem, since the additional CP-odd flavor-invariants felt by the colored particles are the same quark invariants found in the absence of $\psi',\psi'^c$ with additional suppressing factors controlled by the tiny lepton Yukawa couplings. Furthermore $N,N'$ are also not important for what concerns the Strong CP Problem because they only interact with the SM via (CP-conserving) gauge-interactions and (CP-conserving) irrelevant couplings. In general, non-renormalizable operators cannot affect $\bar\theta$ appreciably because CP-violation is super-soft and their contributions are therefore suppressed by powers of $m_\psi^2/(4\pi f_{\rm UV})^2$.

We conclude that our picture satisfies all the low energy requirements spelled out in the introduction, including \eqref{naturalness}, \eqref{ysmall}, and \eqref{coincidence}, and robustly solves the Strong CP Problem.

\subsubsection{Cosmology}
\label{sec:cosmo}

We next study the cosmology of our models. Referring back to Figure \ref{fig:s2scales}, we begin with a discussion of the heavy states beyond the SM, and then proceed to lower masses.

Most of the heavy hadrons are unstable and decay into pions or into SM particles and $\psi,{\psi'},N,N'$ via \eqref{dim6CP}, \eqref{dim6m}, \eqref{decayHad}. Similar considerations apply to the heavy $U(1)$ vector. Due to the axial $U(1)$, however, the baryons $\chi\chi\chi$ are practically stable. Hence, if thermalized, the exotic baryons would have decoupled at $T\sim m_{\rm CP}/25$ and subsequently dominated the expansion rate until very recently. To avoid conflict with the physics of BBN, we assume that the temperature of our Universe has never exceeded
\ba\label{RHT}
T_{\rm RH}\ll\frac{m_{\rm CP}}{25}\sim10^{11-13}~{\rm GeV}.
\ea
This guarantees that baryons were never thermalized and their abundance was always safely within acceptable values. The constraint \eqref{RHT} also serves another purpose. If the Universe was hot enough to go through the CP-violating phase transition we may have ended up generating topological defects via the Kibble-Zurek mechanism. These may then have come to dominate the expansion of the Universe, which is also phenomenologically unacceptable. Thanks to the condition \eqref{RHT}, though, the transition to the CP-violating vacuum occurred before or during inflation, so that any abundance of topological defects would have been diluted to an acceptable amount.

Pions, on the other hand, represent no cosmological hazard. They decay via \eqref{dim6CP} into messengers and SM fermions with lifetimes $\tau_\pi\sim 4\pi/(y^2m_\pi)$, where $y\sim10^{-9}$ was introduced in \eqref{yEFT}. Even if produced abundantly at re-heating, they would have disappeared from the plasma when the early Universe was at $T\sim$ few TeV.

Going further down in mass we encounter the states $\psi,\psi^c,{\psi'},{\psi'}^c,N,N'$. Obviously, since $\psi,\psi^c,{\psi'},{\psi'}^c$ mix with the SM quarks and leptons, they are unstable and decayed very quickly as soon as they decoupled. 

The situation is a bit more complicated for $N,N'$. These particles are cosmologically stable and couple to the plasma only via higher-dimensional operators suppressed by the UV cutoff as well as gauge $U(1)$ interactions. For temperatures satisfying \eqref{RHT} they never thermalized. Under the reasonable assumption that they were not directly produced by the inflaton, a tiny population of $N,N'$ was nevertheless generated at, and soon after, re-heating by the annihilation of nearly thermalized $\psi,\psi^c,{\psi'},{\psi'}^c$. The latter processes are mediated by the effective 4-fermion interaction 
\ba
{\cal L}_{\rm eff}\supset-\frac{1}{f^2}J^A_\mu \left[q_N N^\dagger\bar\sigma^\mu N+q_{N'} N'^\dagger\bar\sigma^\mu N'\right]
\ea
where $J^A_\mu=\sum_iq_i~\Psi_i^\dagger\bar\sigma_\mu\Psi_i$ is the $U(1)$ current of the thermalized charged fermions $\Psi_i=\psi,\psi^c,{\psi'},{\psi'}^c$ and $q_i$ their $U(1)$ charges. We would like to provide a quantitative estimate of the energy density carried by the spectators. In order to do so it is enough to focus on the $N$'s because $N'$ have a much smaller mass and their energy density is suppressed by a factor $m_{N'}/m_N\sim\epsilon\lesssim10^{-5}$ compared to that of $N$.

An approximate estimate of the yield $Y_N=n_N/s$ is obtained via the Boltzmann equation $dY_N/dt=\Gamma_{\rm coll}/s$. If the $\Psi_i$'s are taken to have had a thermal distribution, for simplicity, the collision term is given by
\ba
\Gamma_{\rm coll}=q_N^2\left(\sum_i g_iq_i^2\right)\frac{1}{18\pi^5f^4}\left(\frac{7\pi^4 T^4}{120}\right)^2
\ea
where $g_i$ is the multiplicity of each $\Psi_i$ (helicity excluded). From table \ref{table} we have $q_N^2\sum_i g_iq_i^2=400/81$. Finally, the present-day energy in units of the entropy, ${\rho_N}/{s}=\sum_{I=1}^4 m_{N_I}Y_N$ (there is a family of 4 $N$'s in Table \ref{table}), is approximately 
\ba\label{DMden}
\frac{\rho_N}{s}&\sim&\left.\sum_{I=1}^4 m_{N_I}\frac{\Gamma_{\rm coll}}{Hs}\right|_{T_{\rm RH}}\\\no
&\sim&\frac{\rho_{\rm DM}}{s}~\left(\frac{\sum_{I=1}^4 m_{N_I}}{4~{\rm keV}}\right)\left(\frac{10^{13}~{\rm GeV}}{f}\right)^4\left(\frac{T_{\rm RH}}{10^{11}~{\rm GeV}}\right)^3.
\ea
Since this expression is dominated by the large-$T$ regime, where details of re-heating can be important, our computation should be viewed as a qualitative estimate of the actual density. Nevertheless, the message is clear: in the small $T_{\rm RH}$ regime \eqref{RHT}, where eq. \eqref{DMden} was consistently derived, our spectator sector represents generically a subleading component of dark matter. Yet, with some luck it could be a portion of the missing matter of the Universe, with $N$ a good cold dark matter candidate and $N'$ a negligible component of dark radiation.

\subsubsection{Collider Signatures}
\label{sec:collider}

The particles $\psi,\psi^c,{\psi'},{\psi'}^c$ are subject to collider, electroweak, and flavor constraints. A detailed analysis of the quark mediators can be found in \cite{Valenti:2021rdu}, see also \cite{Cherchiglia:2020kut, Cherchiglia:2021vhe}. The bottom line is that most of parameter space is allowed for masses above the TeV.

The physics of the lepton mediators has not been discussed before in this context, but it is easy to show that these states lead to weaker constraints on the parameters of our scenarios compared to the quark mediators. As for the quark mediators, the $\psi'$-$\ell$ mixing can be removed via a rotation of $(\ell,\psi'^c)$. The massive eigenstate couples to the Higgs and the lepton singlets with coupling that up to a unitary rotation is oriented along the direction of the SM lepton Yukawa coupling $Y_e$, i.e. $Y'\propto Y_e$.

The ${\psi'},{\psi'}^c$ are produced in pairs via Drell-Yan, and then decay into leptons and vector bosons or the Higgs. Current constraints are looser than for quark mediators and are not relevant to our models, where $m_{\psi'}\sim m_\psi$ is the natural expectation. Deviations of the $Z^0$ couplings to leptons are constrained at the permille level. The corresponding bounds are not much stronger than those of the quark mediators, however, because the new coupling $Y'$ is proportional to the SM lepton Yukawa and therefore highly hierarchical. The most significant constraint from flavor-violation comes from the non-observation of $\mu\to e\gamma$ and is well under control for $|Y_e^{-1}Y'|\lesssim300~m_{\psi'}/{\rm TeV}$. CP-violation, including the electric dipole moment of the electron, is strongly suppressed. Overall, we conclude that lepton mediators are allowed to live at the TeV scale.

\subsubsection{Baryogenesis}
\label{sec:baryogenesis}

Explaining the observed baryon asymmetry may at first sight appear difficult in our scenarios, as baryogenesis necessitates of both new sizable CP-violating phases and new interactions with the SM. Yet, successfull baryogenesis above the weak scale does not require new couplings to the {\emph{colored sector}}, and therefore does not immediately jeopardize our solution of the Strong CP Problem. In fact, it may be realized simply adding new physics with CP-violating couplings to the leptons. Low-energy leptogenesis thus appears to be the most natural and safe option in these models. 

A slight modification of our model has all the necessary ingredients. Suppose we replace the spectator sector of Table \ref{table} with this more complicated set of fermions neutral under the SM and the new $SU(3)$ but chiral under the $U(1)$:

\begin{table}[h!]
	\begin{center}
		\begin{tabular}{c||c}
			& $U(1)$    \\
			\hline\hline
			$N_{1,2}$  & $-\frac{1}{3}$  \\
			$N'_{1,\cdots,5}$  & $-\frac{2}{3}$ \\
			$X_{1,2,3}$  & $+\frac{1}{2}$  \\
			$X'_{1,\cdots,5}$ & $-\frac{1}{6}$ 
		\end{tabular}
	\end{center}
\end{table}

The main difference compared to the spectator sector of Table \ref{table} is that with this modified field content one finds a renormalizable coupling $\psi_2'^cHN$ as well as dimension-6 interactions that generate complex Majorana masses of order the TeV for $N,N',X$ (note that $N'$ mixes with $N$). The state $X'$ instead obtains a mass from dimension-9 interactions and is expected to be of order \eqref{N'mass}. In an appropriate portion of the parameter space the modified model may thus generate the observed baryon asymmetry via resonant decays $N,N'\to {{\psi'}^c}^\dagger {H}^\dagger, {\psi'}^c H\to\ell^\dagger{H}^\dagger,\ell H$ as studied in \cite{Pilaftsis:2003gt}. The other fields of the spectator sector would behave qualitatively as in the scenario discussed in Section \ref{sec:model}; $X$ would be a cold dark matter candidate and $X'$ a negligible part of radiation. The crucial difference compared to our earlier model is that here $X$ is much heavier and for this reason has an abundance that is roughly a factor $10^9$ larger compared to \eqref{DMden}. This implies that the re-heating temperature is allowed to be three orders of magnitude smaller, $T_{\rm RH}\sim10^8$ GeV. We can thus have a viable dark matter candidate comfortably within the allowed regime \eqref{RHT}. The phenomenology of this modified version of our scenario is quite rich and would deserve further scrutiny. Our purpose here is merely to demonstrate that there is no structural obstruction to incorporating a mechanism for baryogenesis in our scenarios.

\section{Discussion}
\label{sec:discussion}

Spontaneous CP breaking provides a possible avenue to tackle the Strong CP Problem. Within such a framework CP is postulated to be an exact property of the UV and a non-generic mechanism of mediation between the sector responsible for CP breaking and the SM must be in place. These constructions must be able to explain why CP-violation occurs at parametrically low scales, see \eqref{naturalness}, and why the couplings to the CP-violating sector are small, see \eqref{ysmall}. Most importantly they should account for a remarkable ``coincidence" \eqref{coincidence}: the (CP-conserving) mass $m_\psi$ of the mediators and their (CP-violating) mass-mixing $\xi$ with the SM quarks must be of comparable size. In this paper we demonstrated that it is possible to obtain a class of fully realistic models with all the required properties. We discussed in detail a specific class of models, but our constructions are by no means unique. 

In our scenarios the non-generic structure required to successfully mediate CP-violation follows from gauge invariance, in particular from the gauging of the global $U(1)$ introduced in Section \ref{sec:intro}. Physics at the UV cutoff must be CP-invariant but can be otherwise generic. The Lagrangian we analyzed is consistently the most general one compatible with gauge invariance and the assumed field content. We imposed no internal global symmetry by hand and no structure in the couplings beyond the SM (of course the SM Yukawas are hierarchical, as usual). In our framework CP is spontaneously broken by the vacuum condensate $\langle\Sigma\rangle\sim f$ of an exotic strong dynamics such that the order parameter can be naturally small compared to the cutoff, as required by \eqref{naturalness}. In addition, no fundamental scalar is introduced beyond the SM Higgs doublet. These features ensure that the smallness of $|\bar\theta|$ is achieved truly as the result of a robust dynamical mechanism and not of a hidden fine-tuning of the UV parameters.

The ``coincidence" \eqref{coincidence} is explained because both the mass ($m_\psi$) and the mixing ($\xi$) with the SM arise from interactions involving the mediators and the CP-violating sector. As long as the UV is sufficiently generic, the two couplings are comparable and eq. \eqref{coincidence} follows. This picture requires the $U(1)$ to be chiral, namely $z_{\psi^c}\neq-z_\psi$ in Table \ref{tab:EFT}, with both $m_\psi,\xi$ order parameters ($z_{\psi}\neq0,-z_{\psi^c}$). Crucially, very general theorems guarantee that $m_\psi$ be CP-even while $\xi$ is CP-odd, despite both scales originate from condensates of the CP-violating sector. We argued that this essential feature is a consequence of the breaking of CP being dynamical and of the gauging of the axial $U(1)$. Whether analogous arguments can be put forward in (Supersymmetric) models with fundamental scalars remains an open question.

Besides providing UV completions of the scenarios of Table \ref{tab:EFT} with the fundamental properties \eqref{naturalness}, \eqref{ysmall}, and \eqref{coincidence}, our models also lead to important low energy predictions. For example, the non-observation of the neutron electric dipole moment translates into an upper bound on the masses of the exotic colored mediators $\psi,\psi^c$: 
\ba\label{uppermpsi}
m_\psi\sim4\pi f^3/f_{\rm UV}^2\lesssim{\rm few}~{\cal O}(10)~{\rm TeV}.
\ea
The fermionic messengers, a defining feature of Nelson-Barr models, must therefore be accessible {\emph{directly}} at present and future hadronic colliders and {\emph{indirectly}} in CP- and flavor-violating observables. Moreover, the generic presence of exotic states carrying accidental global symmetries induces a few interesting cosmological signatures. This demonstrates that solutions of the Strong CP Problem via spontaneous CP violation can be very predictive.

There are more lessons we can learn from our UV completions. In our scenarios the interactions between the sector responsible for the spontaneous breaking of CP and the SM are all non-renormalizable and suppressed by a high scale $f_{\rm UV}$ of order the Planck scale. This property, forced upon us by gauge invariance, has two important implications. First, it ensures that the corrections to $\bar\theta$ from the excitations of the CP-breaking sector are negligible. In particular, contributions proportional to the couplings $y$ of Section \ref{sec:intro} are automatically killed, so \eqref{ysmall} follows from the non-renormalizability of the corresponding interaction. Second, it implies the scale of spontaneous CP breaking (here $4\pi f$) and the scale $m_\psi$ at which CP-violation is actually communicated to the SM are decoupled. This hierarchy 
\ba\label{supersoft}
m_\psi\ll4\pi f\ll 4\pi f_{\rm UV}
\ea
makes CP-violation within the SM super-soft. With a super-soft CP-violating scale the effect of possible additional heavy physics decouples from $\bar\theta$. That is, because of \eqref{supersoft} the existence of new physics unrelated to the Strong CP Problem, characterized by masses $m\gg m_\psi$ and sizable couplings to the SM but not to the CP-violating sector, is not severely constrained in our scenarios. Precisely, the impact of heavy particles on $\bar\theta$ decouples as powers of $m^2_\psi/m^2\ll1$. New physics above the TeV may thus safely be invoked to address other puzzles in physics beyond the SM, like the origin of the SM flavor hierarchy, or to {\emph{partially}} stabilize the Weak-Planck scale hierarchy without spoiling our solution of the Strong CP Problem. Said differently, there need not be a desert between $4\pi f$ and the TeV scale in Fig. \ref{fig:s2scales}! Moreover, there is no structural obstruction to realize low-scale leptogenesis either. Specifically, in Section \ref{sec:baryogenesis} we have shown that such a possibility finds a natural implementation in our models.

A large portion of parameter space of our models remains currently compatible with messengers at the TeV scale. Yet, \eqref{uppermpsi} makes us confident that, eventually, experiments will be able to discover or completely exclude this approach to the Strong CP Problem.

\section*{Acknowledgments}

We thank F. D'Eramo for discussions. This project has received support from the EU Horizon 2020 research and innovation programme under the Marie Sklodowska-Curie grant agreement No 860881-HIDDeN.

\bibliography{biblio.bib}

\begin{thebibliography}{10}

\bibitem{Choi:1992xp}
Ki-woon Choi, David~B. Kaplan, and Ann~E. Nelson.
\newblock {Is CP a gauge symmetry?}
\newblock {\em Nucl. Phys. B}, 391:515--530, 1993.
\newblock \href {http://arxiv.org/abs/hep-ph/9205202}
  {\path{arXiv:hep-ph/9205202}}, \href
  {https://doi.org/10.1016/0550-3213(93)90082-Z}
  {\path{doi:10.1016/0550-3213(93)90082-Z}}.

\bibitem{Dine:1992ya}
Michael Dine, Robert~G. Leigh, and Douglas~A. MacIntire.
\newblock {Of CP and other gauge symmetries in string theory}.
\newblock {\em Phys. Rev. Lett.}, 69:2030--2032, 1992.
\newblock \href {http://arxiv.org/abs/hep-th/9205011}
  {\path{arXiv:hep-th/9205011}}, \href
  {https://doi.org/10.1103/PhysRevLett.69.2030}
  {\path{doi:10.1103/PhysRevLett.69.2030}}.

\bibitem{Carpenter:2009zs}
Linda~M. Carpenter, Michael Dine, and Guido Festuccia.
\newblock {Dynamics of the Peccei Quinn Scale}.
\newblock {\em Phys. Rev. D}, 80:125017, 2009.
\newblock \href {http://arxiv.org/abs/0906.1273} {\path{arXiv:0906.1273}},
  \href {https://doi.org/10.1103/PhysRevD.80.125017}
  {\path{doi:10.1103/PhysRevD.80.125017}}.

\bibitem{Dine:2015jga}
Michael Dine and Patrick Draper.
\newblock {Challenges for the Nelson-Barr Mechanism}.
\newblock {\em JHEP}, 08:132, 2015.
\newblock \href {http://arxiv.org/abs/1506.05433} {\path{arXiv:1506.05433}},
  \href {https://doi.org/10.1007/JHEP08(2015)132}
  {\path{doi:10.1007/JHEP08(2015)132}}.

\bibitem{Hiller:2001qg}
Gudrun Hiller and Martin Schmaltz.
\newblock {Solving the Strong CP Problem with Supersymmetry}.
\newblock {\em Phys. Lett. B}, 514:263--268, 2001.
\newblock \href {http://arxiv.org/abs/hep-ph/0105254}
  {\path{arXiv:hep-ph/0105254}}, \href
  {https://doi.org/10.1016/S0370-2693(01)00814-0}
  {\path{doi:10.1016/S0370-2693(01)00814-0}}.

\bibitem{Vecchi:2014hpa}
Luca Vecchi.
\newblock {Spontaneous CP violation and the strong CP problem}.
\newblock {\em JHEP}, 04:149, 2017.
\newblock \href {http://arxiv.org/abs/1412.3805} {\path{arXiv:1412.3805}},
  \href {https://doi.org/10.1007/JHEP04(2017)149}
  {\path{doi:10.1007/JHEP04(2017)149}}.

\bibitem{Nelson:1983zb}
Ann~E. Nelson.
\newblock {Naturally Weak CP Violation}.
\newblock {\em Phys. Lett. B}, 136:387--391, 1984.
\newblock \href {https://doi.org/10.1016/0370-2693(84)92025-2}
  {\path{doi:10.1016/0370-2693(84)92025-2}}.

\bibitem{Barr:1984qx}
Stephen~M. Barr.
\newblock {Solving the Strong CP Problem Without the Peccei-Quinn Symmetry}.
\newblock {\em Phys. Rev. Lett.}, 53:329, 1984.
\newblock \href {https://doi.org/10.1103/PhysRevLett.53.329}
  {\path{doi:10.1103/PhysRevLett.53.329}}.

\bibitem{Valenti:2021rdu}
Alessandro Valenti and Luca Vecchi.
\newblock {The CKM Phase and $\bar\theta$ in Nelson-Barr Models}.
\newblock 5 2021.
\newblock \href {http://arxiv.org/abs/2105.09122} {\path{arXiv:2105.09122}}.

\bibitem{Vafa:1983tf}
C.~Vafa and Edward Witten.
\newblock {Restrictions on Symmetry Breaking in Vector-Like Gauge Theories}.
\newblock {\em Nucl. Phys. B}, 234:173--188, 1984.
\newblock \href {https://doi.org/10.1016/0550-3213(84)90230-X}
  {\path{doi:10.1016/0550-3213(84)90230-X}}.

\bibitem{Hook:2014cda}
Anson Hook.
\newblock {Anomalous solutions to the strong CP problem}.
\newblock {\em Phys. Rev. Lett.}, 114(14):141801, 2015.
\newblock \href {http://arxiv.org/abs/1411.3325} {\path{arXiv:1411.3325}},
  \href {https://doi.org/10.1103/PhysRevLett.114.141801}
  {\path{doi:10.1103/PhysRevLett.114.141801}}.

\bibitem{Raby:1979my}
Stuart Raby, Savas Dimopoulos, and Leonard Susskind.
\newblock {Tumbling Gauge Theories}.
\newblock {\em Nucl. Phys. B}, 169:373--383, 1980.
\newblock \href {https://doi.org/10.1016/0550-3213(80)90093-0}
  {\path{doi:10.1016/0550-3213(80)90093-0}}.

\bibitem{Cherchiglia:2020kut}
A.~L. Cherchiglia and C.~C. Nishi.
\newblock {Consequences of vector-like quarks of Nelson-Barr type}.
\newblock {\em JHEP}, 08:104, 2020.
\newblock \href {http://arxiv.org/abs/2004.11318} {\path{arXiv:2004.11318}},
  \href {https://doi.org/10.1007/JHEP08(2020)104}
  {\path{doi:10.1007/JHEP08(2020)104}}.

\bibitem{Cherchiglia:2021vhe}
A.~L. Cherchiglia, G.~De~Conto, and C.~C. Nishi.
\newblock {Flavor constraints for a Vector-like quark of Nelson-Barr type}.
\newblock 3 2021.
\newblock \href {http://arxiv.org/abs/2103.04798} {\path{arXiv:2103.04798}}.

\bibitem{Pilaftsis:2003gt}
Apostolos Pilaftsis and Thomas E.~J. Underwood.
\newblock {Resonant leptogenesis}.
\newblock {\em Nucl. Phys. B}, 692:303--345, 2004.
\newblock \href {http://arxiv.org/abs/hep-ph/0309342}
  {\path{arXiv:hep-ph/0309342}}, \href
  {https://doi.org/10.1016/j.nuclphysb.2004.05.029}
  {\path{doi:10.1016/j.nuclphysb.2004.05.029}}.

\end{thebibliography}

\bibliographystyle{unsrturl}

\end{document}